\documentclass{elsart}
\usepackage{graphicx,epsfig,amssymb}
\newcommand\textrmn[1]{\textrm{\scriptsize #1}}
\journal{Physics Letters B}
\begin{document}
\begin{frontmatter}

\title{Search for Rare and Forbidden 3-body Di-muon Decays of the Charmed Mesons
$D^+$ and $D_s^+$}

The FOCUS Collaboration\footnotemark

\author[ucd]{J.~M.~Link}
\author[ucd]{P.~M.~Yager}
\author[cbpf]{J.~C.~Anjos}
\author[cbpf]{I.~Bediaga}
\author[cbpf]{C.~G\"obel}
\author[cbpf]{J.~Magnin}
\author[cbpf]{A.~Massafferri}
\author[cbpf]{J.~M.~de~Miranda}
\author[cbpf]{I.~M.~Pepe}
\author[cbpf]{E.~Polycarpo}   
% ...

\address[ucd]{University of California, Davis, CA 95616}
\address[cbpf]{Centro Brasileiro de Pesquisas F\'isicas, Rio de Janeiro, RJ, Brasil}
\address[cinv]{CINVESTAV, 07000 M\'exico City, DF, Mexico}
\address[cu]{University of Colorado, Boulder, CO 80309}
\address[fnal]{Fermi National Accelerator Laboratory, Batavia, IL 60510}
\address[fras]{Laboratori Nazionali di Frascati dell'INFN, Frascati, Italy I-00044}
\address[ugj]{University of Guanajuato, 37150 Leon, Guanajuato, Mexico} 
\address[ui]{University of Illinois, Urbana-Champaign, IL 61801}
\address[iu]{Indiana University, Bloomington, IN 47405}
\address[korea]{Korea University, Seoul, Korea 136-701}
\address[kp]{Kyungpook National University, Taegu, Korea 702-701}
\address[milan]{INFN and University of Milano, Milano, Italy}
\address[nc]{University of North Carolina, Asheville, NC 28804}
\address[pavia]{Dipartimento di Fisica Nucleare e Teorica and INFN, Pavia, Italy}
\address[pr]{University of Puerto Rico, Mayaguez, PR 00681}
\address[sc]{University of South Carolina, Columbia, SC 29208}
\address[ut]{University of Tennessee, Knoxville, TN 37996}
\address[vu]{Vanderbilt University, Nashville, TN 37235}
\address[wisc]{University of Wisconsin, Madison, WI 53706}
\addtocounter{footnote}{-1}
\footnotetext{\mbox{See http://www-focus.fnal.gov/authors.html for additional author information.}}

\begin{abstract}
Using a high statistics sample of photo-produced charm particles from the FOCUS
experiment at Fermilab, we report results of a search for eight rare 
and Standard-Model-forbidden decays:  
$D^+,~D_{s}^+ \to h^\pm\mu^\mp\mu^+$ (with $h=\pi, K$).
Improvement over previous results by a factor of 1.7--14 is realized. 
Our branching ratio upper limit $D^+\to \pi^+\mu^-\mu^+$ of $8.8\times10^{-6}$
at the 90\% C.L. is below the current MSSM R-Parity violating constraint.
\end{abstract}
\end{frontmatter}


%
%
\section{Introduction}

The search for rare and forbidden decays of charm particles is enticing since
Standard Model (SM) predictions for interesting decays tend to be beyond the reach of current
experiments, and a signal is an indication of unexpected physics.

Standard Model predictions \cite{Burdman:2001tf,Fajfer:2001sa,Singer:1996it} for the branching 
ratios of rare
decays $D^+, D_s^+ \rightarrow h^+ \mu^- \mu^+$ (with $h = \pi, K$) are
dominated by long range effects which are notoriously difficult to
calculate.  Even so, the differences between the three predictions in
\cite{Burdman:2001tf,Fajfer:2001sa,Singer:1996it}, although using individual 
treatments for the decay spectra, are quite manageable.  For example, the 
predicted integrated rate for $D^+\rightarrow \pi^+ \mu^- \mu^+$ varies by only a 
factor of 2 while the experimental limits are a factor of 5--10 away. 
Minimal Supersymmetric  Standard Model (MSSM) R-Parity violating extensions can significantly 
increase this rate. Experimental results for $D^+\to \pi^+\mu^-\mu^+$ currently set the
best constraint for the product of the MSSM R-Parity violating couplings:
$|\lambda_{22k}'\lambda_{21k}|'<0.004$ \cite{Burdman:2001tf}. Until experiments 
reach the SM limit for these rare decays,
a signal indicates new physics or a needed refinement in the interpretation of
the SM. Decays of the form $D^+,~D_{s}^+ \to h^-\mu^+\mu^+$ (with $h=\pi, K$) 
are forbidden in the SM since they violate lepton number conservation, and a signal
in these modes is direct evidence of new physics.

In this paper, we present new upper limits for the branching ratios of 3-body di-muonic
decays of the $D^+$ and $D_s^+$ mesons mentioned above. Unless specifically mentioned,
all results include a lower limit of 0 at the 90\% C.L..
Our results represent a factor of 1.7--14 improvement over previous
experimental limits \cite{Frabetti:1997wp,Aitala:1999db}. The result for the branching ratio
upper limits of $D^+ \to \pi^+ \mu^+ \mu^-$ of $8.8 \times 10^{-6} ~\char'100$ 90\% C.L.
and $D_s^+ \to \pi^+ \mu^+ \mu^-$ of $2.6 \times 10^{-5} ~\char'100$ 90\% C.L. are
both within a factor
of 5 of the Standard Model (long range) predictions  
$1.9\times 10^{-6}$ \cite{Burdman:2001tf} and $6.1\times 10^{-6}$ \cite{Fajfer:2001sa} respectively.  
The 
result for the branching ratio upper limit $D^+ \to \pi^+ \mu^+ \mu^- $ of $8.8 \times
10^{-6}~\char'100$ 90\% C.L. is below the MSSM R-Parity violating constraint \cite{Burdman:2001tf}. 

The data for this analysis
were collected using the Wideband photoproduction experiment FOCUS during the 
1996--1997 fixed-target run at Fermilab. The FOCUS detector is a large aperture, 
fixed-target spectrometer with excellent vertexing and particle identification
used to measure the interactions of high energy photons on a segmented BeO 
target. The FOCUS beamline \cite{Frabetti:1992bn} and 
detector \cite{Frabetti:1990au,Link:2001pg,Link:2002zg,Link:2002ev} have been described elsewhere.

\section{Event Selection}

We look for $D$'s through the
3-body decay chains $D^{+},~D_s^{+}\to h^\mp\mu^\pm\mu^+$ (where the
$h$ represents a pion or a kaon) for rare decays and $D^{+}\to K^-\pi^+\pi^+$ 
or $D_s^{+}\to K^-K^+\pi^+$ for normalization (charge conjugate modes are
implied throughout this paper). 
In order to search for the set of cuts that provides
signal optimization, we place initial (loose) requirements on
the reconstructed data to produce a base sample, and 
then we place a series of (tighter) cuts on the base sample.
The loose requirements consist of acceptance, momentum,  vertexing, and
particle identification cuts. Note that for all cuts, care is taken to 
ensure that the normalization modes receive the same cuts as the di-muon modes where possible.

Due to the finite lifetime and Lorentz boost
of charm candidates, the primary interaction vertex and secondary decay
vertex can have a significant separation along the beam direction. Secondary
vertices are formed from 3 candidate tracks, and the resulting momentum vector 
of the 3 tracks is used as a seed to search for a primary vertex \cite{Frabetti:1990au}.
We require that the confidence level of the secondary vertex fit (DCL) exceed 1\%,
the confidence level of the primary vertex fit exceed 1\%,
the significance of separation between the primary and secondary 
vertices ($\ell/\sigma_\ell$) exceed 5, and the confidence level that any of the
secondary tracks is consistent with the primary vertex (ISO1) be less than
10\%. The latter cut is included to remove the contamination from $D^{*+}$
decays and other tracks originating from the primary that could be
confused with secondary tracks.

We use the \v{C}erenkov system \cite{Link:2001pg} to identify pions and
kaons. 
For each track, $W_\textrmn{obs} = - 2 \log {\mathcal{(L)}}$ is computed, where
$\mathcal{L}$ is the likelihood that a track is consistent with a given
particle hypothesis. For a candidate kaon, we require
$W_\textrmn{obs}(\pi)-W_\textrmn{obs}(K)$ (kaonicity) be greater than 0.5.
For a candidate pion in a rare mode, we require
$W_\textrmn{obs}(K)-W_\textrmn{obs}(\pi)$ (pionicity) be greater than -15.

Muon candidates are required to be within the acceptance of either of
the 2 muon systems in FOCUS \cite{Link:2002ev}.
We require the tracks in a normalizing mode corresponding to muon 
tracks in a rare decay mode be in the acceptance
of one of the muon systems as well. Since the rate of muon misidentification
increases at low momentum, we place a requirement that the momentum of ``muon''
tracks within the acceptance of the wide angle (outer) muon system ($P_\mu^\textrmn{~outer}$) be
greater than $6~\textrm{GeV/c}$ and those within the low angle (inner) muon system ($P_\mu^\textrmn{~inner}$)
be greater than $8~\textrm{GeV/c}$. All muon candidates are required to have associated hits in 
either muon system sufficient to meet a minimum confidence level, $\mu_{CL}$, 
of 1\% for the muon hypothesis, and must pass additional muon cuts
depending on the system traversed.
For the outer muon system, muon candidates must traverse a minimum of
150 cm of material, produce hits in 2 of 3 planes of the detector,
and these 2 (or more) hits (called a cluster) must not be shared by the 
other muon candidate. For the inner system, at least 4 of 6 planes of the 
detector must record hits consistent with the candidate track,
no more than 2 of these hits can be shared with the other candidate muon track,
and a fit to the hypothesis that the inner muon candidate track had 
the same measured momentum in both magnets traversed was required to exceed 1\%.
This last cut is used to reduce contamination from particles that decay 
and produce a muon as they traverse the spectrometer. This cut is also
applied to the pions in the normalization modes, and the lowest momentum
kaon, when possible, for the $D_s^+$ normalization mode. Finally,
all other tracks in the event are fit to the muon hypothesis using the 
hits from a candidate muon, and the highest confidence level from the fits, 
ISO$_{\mu}$, is saved. No ISO$_{\mu}$ cut was 
required for the base sample, but a $10\%$ ISO$_{\mu}$ is included in the 
set of cuts used for sensitivity optimization.

For the $D_s^+$ normalization signal, $D_s^+\to K^+K^-\pi^+$, a 
cut was applied to reduce the reflection when one
of the pions from $D^+\to K^-\pi^+\pi^+$ is misidentified as a kaon. 
Under the hypothesis that the same sign $D_s^+$ kaon track is assumed
to be pion, the invariant mass is calculated. If the new invariant mass 
is within 3 standard deviations
of the $D^+$ mass, the kaonicity of the same sign kaon must exceed 6.
Additionally, we require that the reconstructed $K^+K^-$ invariant mass be
within 3 standard deviations of the $\phi$ mass.

Our analysis methodology (Section 3) requires a base sample of events 
of sufficiently small size (150 events) for a reasonable processing 
time. Base samples were obtained by applying the minimum $\ell/\sigma_\ell$ cut in 
the range of 5 to 8 which brought the sample size below 150.  
The series 
of cuts applied to these samples was arranged into a grid.  This cut 
grid was based on kinematic variables, particle ID algorithm results,
vertex quality, and event topology. For example, once the lower cut 
in $\ell/\sigma_\ell$ was determined, the $\ell/\sigma_\ell$ cut was allowed 
to vary in steps of 2 units up to a maximum of 21 or 22. 
The cuts used have been shown to be 
effective for other charm decays besides those presented in this 
analysis.  

\begin{figure}[htb]
\begin{center}
\epsfig{file=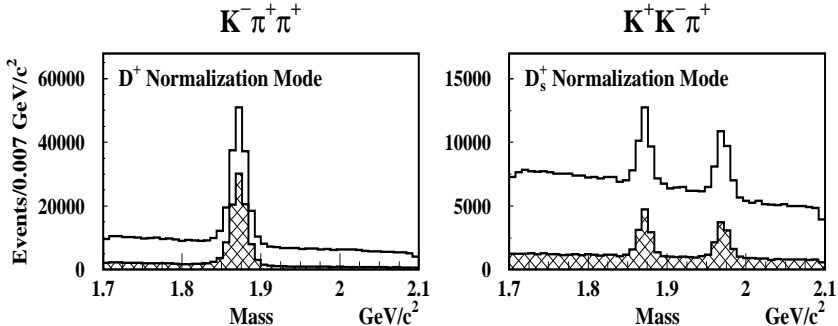,height=5cm,width=12cm}
\caption{Modes used to normalize the rare decay modes. The solid histograms
represent the loosest cuts employed in the
analysis, while the superimposed cross-hatched histograms represent
data which has had the tightest cuts used in the analysis applied.
Notice the large reduction in background relative to the signal 
for both the $D^+$ and the $D_s^+$ modes over the range of cuts used.}
\label{fig:norm}
\end{center}
\end{figure}

Since the kaonicity cut can be applied to the kaon modes identically, there
is an inherent increase in the size of the cut grid when the pionicity cut is
used for the pion modes. To keep the size of the cut grid roughly the same,
and to prevent memory overflows in software that was difficult to modify,
we reduced the size of the cut grid for the pion modes. The grid (see Table \ref{tab:cuts}) includes 
cuts on $\ell/\sigma_\ell$, ISO1, DCL, kaonicity, pionicity, $\mu_{CL}$, ISO$_{\mu}$, and the momentum of muon candidates.
The normalizing modes used to compute the branching ratios for $D^{+}$ and $D_s^{+}$
are shown in Figure \ref{fig:norm} for the loosest cuts in the grid.

\begin{table}[htb]
\caption{Analysis cuts used in the cut grid. Variables are described in the text. The
best cut on average represents a point on the cut grid used in a systematic
check of our result that is described in the ``Systematic Checks and Results'' section
of the paper. Cuts indicated by $\{~\}$ are applied only to the kaon modes to
keep the cut grid about the same size for kaon and pion modes. Notice that
the cuts removed for the pion modes are chosen either very far from the 
``Best Cut'' (explained later in the text), or represent a small reduction in the
stepping of a cut that varies logarithmically (ISO1).
}
\label{tab:cuts}
\begin{center}
\begin{tabular}{cccccl} \hline 
 Variable              &  Cut Values Used in the Grid             & Best Cut on Average          \\ \hline       
$\ell/\sigma_\ell$        & $>5\to 22$            &  $>13~(D^+),~10~(D_s^+)$     \\ \hline
ISO1            	  & $<0.1,\{0.03\},0.01,\{0.003\},0.001$         &  $<0.1$                      \\ \hline
DCL	                  & $>1\%,2\%,4\%$                        &  $>1\%~(D^+),~2\%~(D_s^+)$   \\ \hline
Kaonicity (kaon modes)    & $>0.5,1.0,2.0$                        &  $>1.0$                      \\ \hline
Pionicity (pion modes)    & $>-15,-3,-1$                          &  $>-3$                      \\ \hline
$\mu_{CL}$                 &  $>1\%,5\%,10\%$                      &  $>5\%$                      \\ \hline
ISO$_{\mu}$    & $<0.10,1.0$                          &  $<1.0$                     \\ \hline 
P$_\mu^\textrmn{~inner}~(\textrm{GeV/c})$   &  $>8,9,10,11,12,\{14\}$ &  $>9 $			 \\ \hline
P$_\mu^\textrmn{~outer}~(\textrm{GeV/c})$  &  $>6,7,8,9,\{10\}$	 &  $>7 $			\\ \hline
\end{tabular}
\end{center}
\end{table}

In order to perform signal optimizations, calculate efficiencies, and estimate yields, 
there is a distinction
made between the signal region, where events for the desired mode
are expected to occur, and the background sidebands, where the amount
of signal is expected to be minimal. The background sidebands are used
to estimate the amount of contamination in the signal region. 
We define the signal region to be within 2 
standard deviations of the
nominal reconstructed parent particle mass (i.e. either the $D^+~(\pm~20~{\rm{MeV/c^2}})$ 
or the $D_s^+~(\pm~18~{\rm{MeV/c^2}})$)
and the background region to be
any invariant mass reconstructed outside the signal regions 
between 1.7 and 2.1 ${\rm{GeV/c^2}}$. 
An exception is made for $D^+,~D_{s}^+\to K^-\mu^+\mu^+$.
For $D^+\to K^-\mu^+\mu^+$ the background sidebands are required to 
be between 1.75 and 2.1 ${\rm{GeV/c^2}}$. This approximately splits the 
expected contribution of $D^+\to K^-\pi^+\pi^+$, where the 2 pions are both
mis--identified as muons, equally between signal and background regions.
For $D_{s}^+\to K^-\mu^+\mu^+$ we require the lower sideband to begin 2 standard
deviations above the $D^+$ mass. This effectively removes the  $D^+\to K^-\pi^+\pi^+$
signal that comes from mis--identified muons. The centroid and standard deviation for each mode
are determined by fitting the reconstructed parent 
particle mass in the normalization mode. Since there is a small shift in the 
mass centroid between data and Monte Carlo, we use the data to determine the
regions for data and Monte Carlo to determine the regions for Monte Carlo.

The shape of the background in each rare mode is determined using a large sample of 
photoproduction Monte Carlo events where all charm species and known decay modes 
are simulated.
The shape of the background is used to determine $\tau$, which is the  
ratio of the number of background events expected in the sideband region to
the number of background events expected in the signal region.
An average $\tau$ is computed for each mode from all the $\tau$'s in the 
cut grid for that particular mode. We find that the ratio of Monte Carlo
efficiencies and $\tau$ are stable over the cut combinations in the grid, but 
we require at 
least 10 surviving Monte Carlo events to determine 
each $\tau$ used in the average to prevent variations due to low
Monte Carlo statistics in the signal region. 

These definitions are used in the determination of the 
branching ratio limits in the analysis described below.

\section{Analysis}

The analysis technique emphasized a careful approach to the
treatment of backgrounds in a limited statistics analysis. The
``blind'' approach, to select cuts that optimize signal 
efficiency relative
to background sidebands, may still lead to a downward 
fluctuation of the sidebands relative to the masked off ``signal'' region
and a more conservative limit on average \cite{Rolke:2002ix}. 
Further, authors 
frequently use the technique outlined
by Feldman and Cousins \cite{Feldman:1997qc} to calculate the confidence levels used in the
calculation of their limits. The Feldman--Cousins approach does not explicitly
include fluctuations of the background. Indeed, the Particle Data Group \cite{James:sr}
suggests 
presenting a measure of the experimental sensitivity, defined as the
average 90\% confidence upper limit of an ensemble of
experiments with the expected background and no true signal, in addition to the 
reported limit whenever experiments quote a result. None of the 
methods suggested in the PDG, including the Cousins--Highland method 
for including systematic errors in upper limits \cite{Cousins:1991qz}, properly 
deal with fluctuations in the background and bias in selecting the data.

For this analysis, we chose a method suggested by Rolke and Lopez \cite{Rolke:2000ij} 
which includes 
the background fluctuations directly into the calculation of the likelihood.
The
composite Poisson probability of finding $x$ events in a signal region and
$y$ events in background sidebands given a signal $\mu$ and a background $b$ is:

\begin{equation}\label{rlke1}
P_{\mu,b}(x,y)={ {(\mu+b)^xe^{-(\mu+b)}} \over{x!} }~
{ {(\tau b)^ye^{-(\tau b)}}\over{y!} }
\end{equation}

where $\tau$ is the expected ratio of the number of background events in the
sideband regions to the signal region. Rolke and Lopez have shown that including
the second Poisson factor in this expression leads to confidence intervals
with better coverage than those of Feldman--Cousins who only consider the
first factor.

In a second paper Rolke and Lopez have shown that bias can also be introduced
during the selection of optimal cuts \cite{Rolke:2002ix}. If a
single cut is chosen based partly on the level of
background in the sidebands (as is typical), there
is a tendency to optimize on downward fluctuations and, hence, to 
underestimate the background level in the signal region. The resultant 
limits from such an ``optimized'' analysis, even though carried out
in a ``blind'' fashion, will not have the correct coverage.

In order to reduce the bias due to selection, Rolke and Lopez suggest
the data be sampled using a ``Dual Bootstrap'' method. In a bootstrap, 
the experimentally observed data set of N events is used to create
an ensemble of many different N-event experiments or data sets,
obtained by random sampling of the original data set allowing repeated events.
In the Dual Bootstrap, two independently bootstrapped
data sets are created. One set is used to optimize the cuts which 
are then applied to the second set in order to calculate the 
confidence intervals. This procedure is repeated 10,000 times and the 
median value for the limits is the final result.  The two bootstrap data sets 
are sufficiently independent that the background estimate from the second 
is very nearly unbiased.

\begin{figure}[htb]
\begin{center}
\epsfig{file=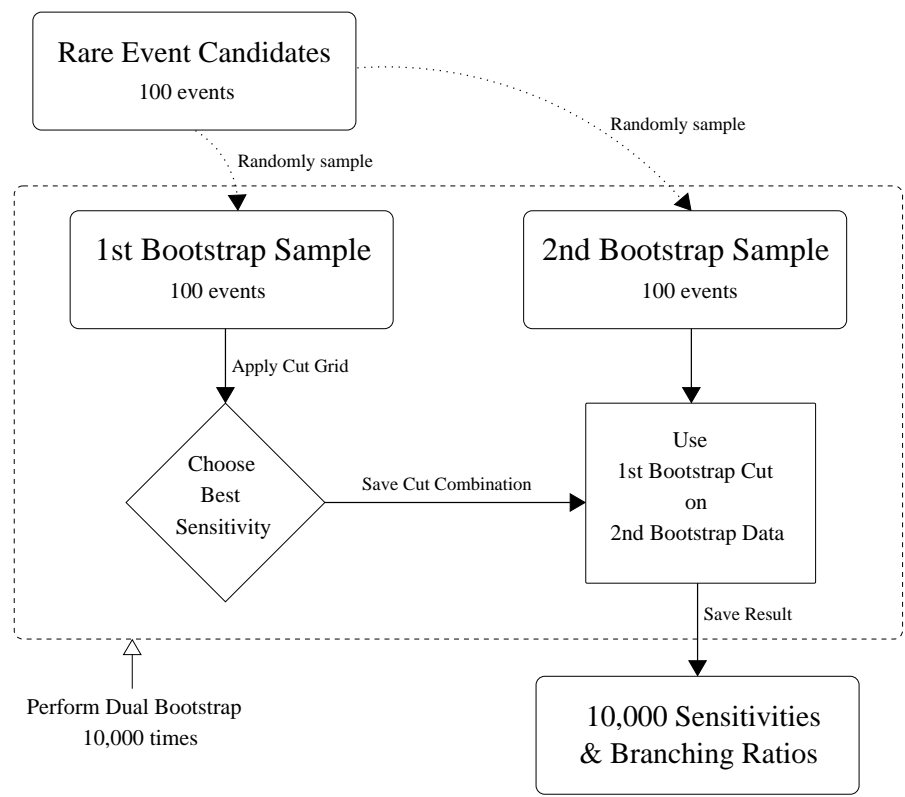,width=12cm}
\caption{A flowchart for the Dual Bootstrap on a sample size of 100 events. 
Note the parallel structure of the sampling with replacement that separates
the cut selection that optimizes sensitivity from the calculation of
the sensitivities and branching ratios used to determine the final result.
Thus, only the data sidebands and the expectation of the shape of the
background is used in the cut selection, even though events may be chosen
from throughout the data set during the bootstrap process.}
\label{fig:bootfig}
\end{center}
\end{figure}

We use the experimental sensitivity 
(see Eqn. \ref{rlke2} and Eqn. \ref{rlke3}) as our figure of
merit to optimize cuts. A matrix or grid of possible data quality 
selection cuts are applied to the first bootstrap data set.  The point 
in the multi-dimensional cut grid which has the best (smallest) 
sensitivity is applied to the second bootstrap data set. 
Limits for the branching ratio and a new
sensitivity are computed from this second set.  The Dual Dootstrap
procedure is shown schematically in Figure \ref{fig:bootfig}.
  
For a given $\tau$
with $y$ sideband events, the average 90\% confidence
upper limit for the number of events in  the signal region when there 
is no true signal, $S_{\tau}(y)$, can be calculated from the 
Rolke-Lopez \cite{Rolke:2000ij} 90\% upper limit table as:

\begin{equation}\label{rlke2}
S_{\tau}(y) = \sum_{x=0}^{\infty} U_{\tau}(x,y) 
        \cdot P_{y/\tau}(x)
\end{equation}

where $U_{\tau}(x,y)$ is the upper
limit of the signal, and $P_{y/\tau}(x)$ is the Poisson probability of
$x$ when the expected background is $y/\tau$. 

The experimental sensitivity is:

\begin{equation}\label{rlke3}
\textrm{Sensitivity} = \textrm{BR}_\textrmn{norm}{{S_{\tau}(y)}\over{Y_\textrmn{norm}}}\epsilon
\end{equation}

where the branching ratio of the normalization mode, $\textrm{BR}_\textrmn{norm}$, 
comes from the PDG \cite{Hagiwara:fs}, $Y_\textrmn{norm}$ is the yield of the 
normalization mode (determined
from a sideband subtraction), and $\epsilon$ is the ratio, determined using Monte Carlo,
of the normalization mode efficiency divided by the rare decay mode efficiency.
For different bootstrap data sets and different cut sets, $y$, $\tau$, $Y_\textrmn{norm}$,
$S_{\tau}(y)$, and $\epsilon$ can be different.  The sensitivity 
meets the requirement of a blind analysis, i.e. it does not depend on 
the number of events observed in the signal region. In a Dual Bootstrap 
procedure one sensitivity is calculated as the best sensitivity for the first data set
and a second (less-biased) sensitivity is calculated when the ``best''
cuts are applied to the second data set.

The 90\% confidence upper limit for the rare branching ratio is:

\begin{equation}\label{rlke4}
\textrm{Upper~Limit} = \textrm{BR}_\textrmn{{norm}}{{Y_\textrmn{rare}}\over{Y_\textrmn{norm}}}\epsilon
\end{equation}

where  $Y_\textrmn{rare}= U_{\tau}(z,y)$ is the Rolke-Lopez 90\% confidence upper limit for the
signal yield given $z$ events in the signal region.  The lower limit has a similar expression.

\begin{figure}[htb]
\begin{center}
\epsfig{file=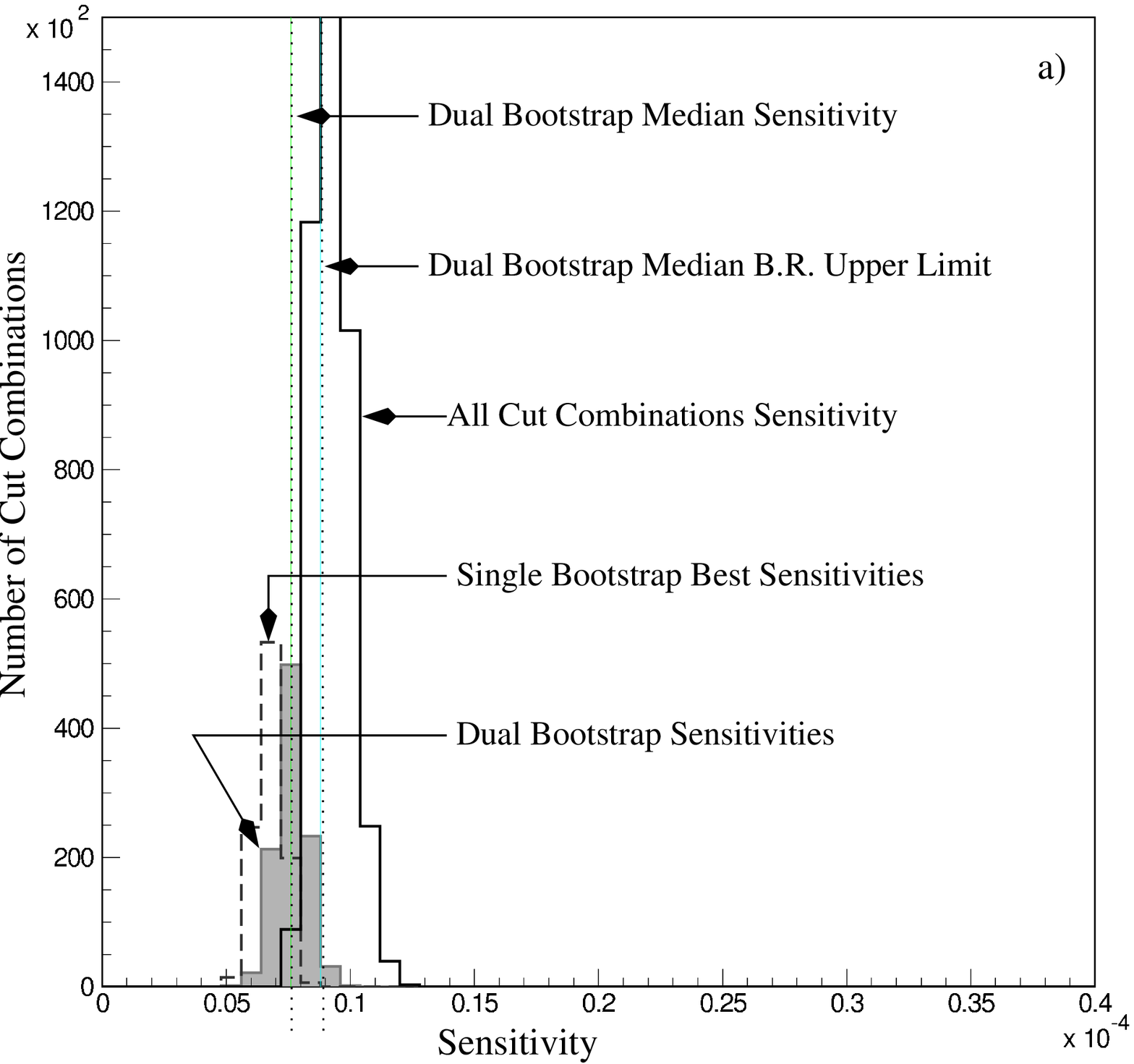,width=6.9cm}\epsfig{file=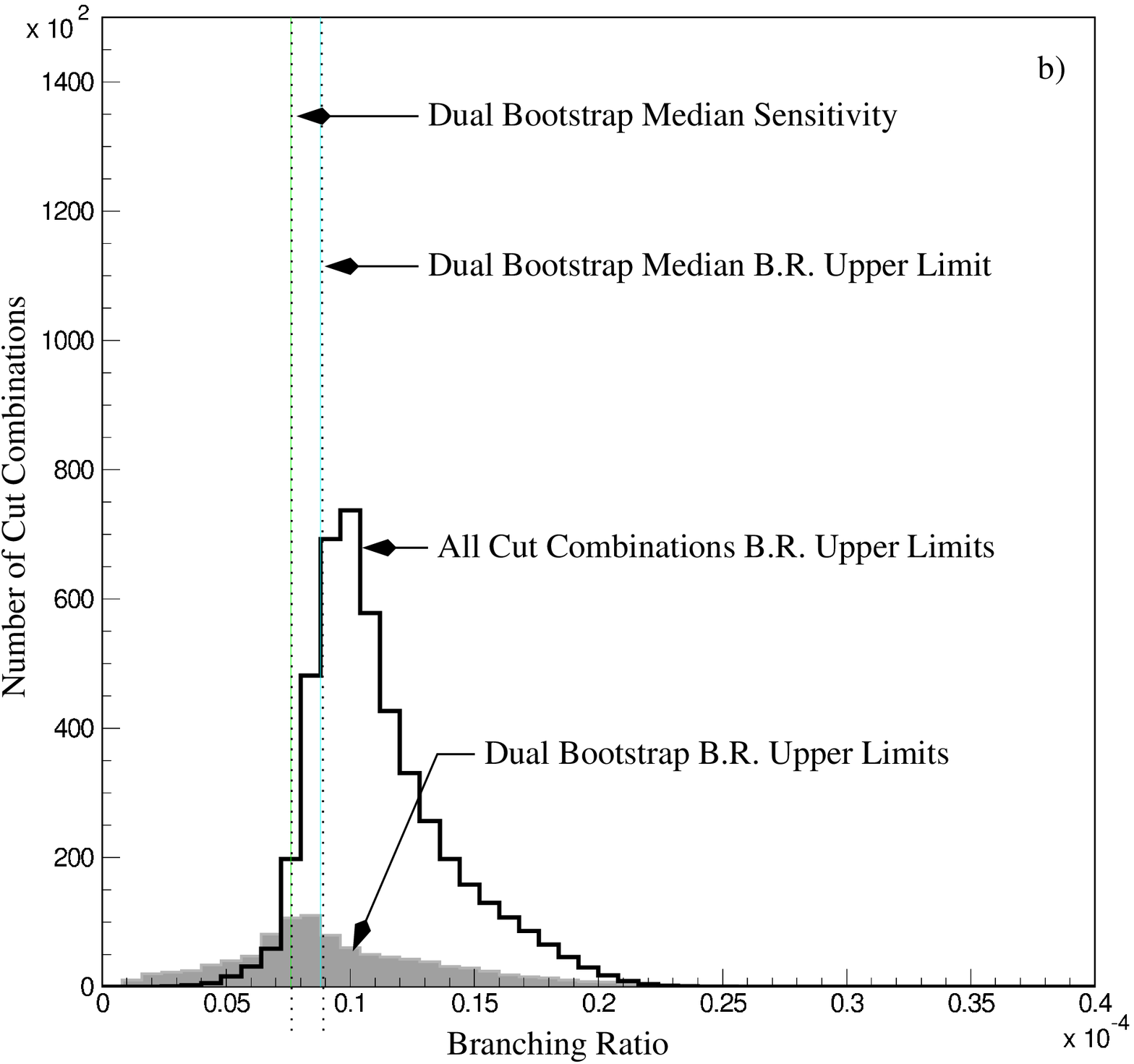,width=6.9cm}
\caption{The distribution of results used in the determination
of the branching ratio confidence interval for the decay mode
$D^+\to\pi^+\mu^-\mu^+$. In plot a), the sensitivity distributions
are shown, and in plot b) the branching ratio distributions are shown.
}
\label{fig:boot}
\end{center}
\end{figure}

The salient features of parameters coming from the Dual Bootstrap 
analysis are illustrated in Figure \ref{fig:boot} for the decay mode $D^+\to\pi^+\mu^-\mu^+$. Several features 
are worth mentioning. The spread of the branching ratio upper limits is in general
larger than the spread of the best sensitivities.
This happens since the spread of the branching ratio upper limits
is dominated by fluctuations in the signal region while the
spread in the best sensitivities stems from the larger
background region used in the estimate. This is a distinct
advantage when determining $\tau$ with large sidebands rather than smaller
sidebands where the fluctuations in the background can become more
problematic. The single bootstrap sensitivities sample the very
lowest end of the sensitivity spectra, and fluctuations often 
place the best bootstrap sensitivity below any calculated from
the data directly. Notice though, that the second bootstrap sensitivities
remove the bias and give a median sensitivity somewhat above
the minimum expected by the data. This is a safeguard against
choosing a single cut that produces an outlier or poor estimation
of the true sensitivity. Thus, the Dual Bootstrap method allows us to
optimize the sensitivity for {\it{each}} decay mode while retaining
correct coverage. Also notice that the Dual Bootstrap branching ratio
upper limits
can cover a larger spread than the original branching ratio upper limits shown
for all cut combinations. This can be understood if one realizes that
for a small number of events in the signal region, the background
can vary considerably in the second bootstrap, and the confidence 
interval calculation amplifies the effect, producing a larger spread.  
 
\section{Systematic Checks and Results}

The largest sources of systematic uncertainty in this analysis are
estimated to come from
uncertainties in the Monte Carlo simulation and uncertainty in the
branching ratios used for the normalizing modes. 
One source of systematic 
uncertainty is included specifically in the upper limits through
the $\tau$ parameter, while the other sources of systematic uncertainty
were included using the technique outlined in \cite{Cousins:1991qz}. Using
this method, the increase, $\Delta U_n$, in the Poisson upper limit 
on the estimate for the number of rare decay events is:

\begin{equation}\label{rlke5}
 \Delta U_n = \frac{1}{2} ~ U_{\!_\textrmn{RL}}^2 ~ \sigma_r^2 ~ 
  \frac{~U_{\!_\textrmn{RL}}+b-n~}{U_{\!_\textrmn{RL}}+b}
\end{equation}

where $U_{\!_\textrmn{RL}}$ represents the Rolke-Lopez 90\% confidence upper limit for the
mean number of signal events, $Y_\textrmn{rare}$, $b$ is the predicted background in the signal region,
$n$ is the number of events found in the signal region and the
total relative systematic uncertainty is $\sigma_r$. For example, uncertainty in the
normalizing branching ratios and efficiencies used in the calculation of
the rare branching ratios is translated to a percent or relative error in the
estimation of $Y_\textrmn{rare}$.
 
The relative systematic uncertainty from the normalizing branching ratios, $\sigma_\textrmn{PDG}$, comes
from the PDG \cite{Hagiwara:fs}. The relative systematic uncertainty stemming from the 
simulation of the data comes from the simulation of the experimental trigger,
$\sigma_\textrmn{Trigger}$,
and the estimation of the efficiency of the outer muon system, $\sigma_{\mu \textrmn{ID}}$. Since the
FOCUS trigger requires a minimum energy be deposited in the 
calorimetry, muons and hadrons will deposit very different energies, and
the trigger simulation must account for any difference. The difference between
three very different
simulations is used to estimate this uncertainty: a full GEANT \cite{Geant:1993xx} simulation of
the calorimetry, a pre--stored shower library generated with GEANT which selects the
calorimetry response based on particle types, energies, and locations, and 
a simulation based on the parameterized average response in data of the calorimetry
based on incident particle types and energies. The relative systematic error
due to the outer muon identification is estimated by looking at the difference
in the Monte Carlo rare decay efficiency for 2 different estimations of the
the outer muon identification efficiency. One method employs the
overlap between the inner and outer systems (very parallel muons coming from 
far upstream of the experiment can impact both systems), and the other method
uses 2 hits in the outer system to predict a third hit. The final source of
systematic error is due to the uncertainty in the modelling of the muon
misidentification used when $\tau$ is calculated for the $D^+\to K^-\mu^+\mu^+$
decay mode. This uncertainty is estimated by boosting the contribution
of $D^+\to K^-\pi^+\pi^+$ in the photoproduction Monte Carlo by twice the
amount needed to match the amount of $D^+\to K^-\pi^+\pi^+$ seen when
one of the pions is misidentified by a muon. The more conservative $\tau$
is then used. The small 
statistical errors from the ratio of Monte Carlo efficiencies and the 
error in the yield of the normalization modes did not contribute significantly 
to the systematic error. 

The sources of relative systematic error for each mode are shown in Table \ref{tab:sys}.
The total relative systematic uncertainty is obtained by adding all the contributions
in quadrature. The effect on the rare branching ratio is calculated for each
bootstrap sample and is naturally included in the ensemble result. 

\begin{table}[htb]
\caption{Contributions to the relative systematic uncertainty, $\sigma_r$, in \%.}
\label{tab:sys}
\begin{center}
\begin{tabular}{cccccl} \hline 
Decay Mode                     &  $\sigma_\textrmn{Trigger}$ & $\sigma_{\mu \textrmn{ID}}$ & $\sigma_\textrmn{PDG}$ & $\sigma_r$ \\ \hline
$D^+ \to K^+ \mu^+ \mu^- $     &  2.8                & 1.9 & 6.7 & 7.5 \\ \hline
$D^+ \to K^- \mu^+ \mu^+ $     &  2.7                & 2.6 & 6.7 & 7.7 \\ \hline
$D^+ \to \pi^+ \mu^+ \mu^- $   &  2.5                & 2.7 & 6.7 & 7.6 \\ \hline
$D^+ \to \pi^- \mu^+ \mu^+ $   &  2.0                & 2.6 & 6.7 & 7.5 \\ \hline
$D_s^+ \to K^+ \mu^+ \mu^- $   &  3.0                & 1.9 & 27.3 & 27.5 \\ \hline
$D_s^+ \to K^- \mu^+ \mu^+ $   &  2.3                & 2.5 & 27.3 & 27.5 \\ \hline
$D_s^+ \to \pi^+ \mu^+ \mu^- $ &  3.6                & 2.7 & 27.3 & 27.7 \\ \hline
$D_s^+ \to \pi^- \mu^+ \mu^+ $ &  3.0                & 2.8 & 27.3 & 27.6 \\ \hline
\end{tabular}
\end{center}
\end{table}

\begin{figure}[htb]
\begin{center}
\epsfig{file=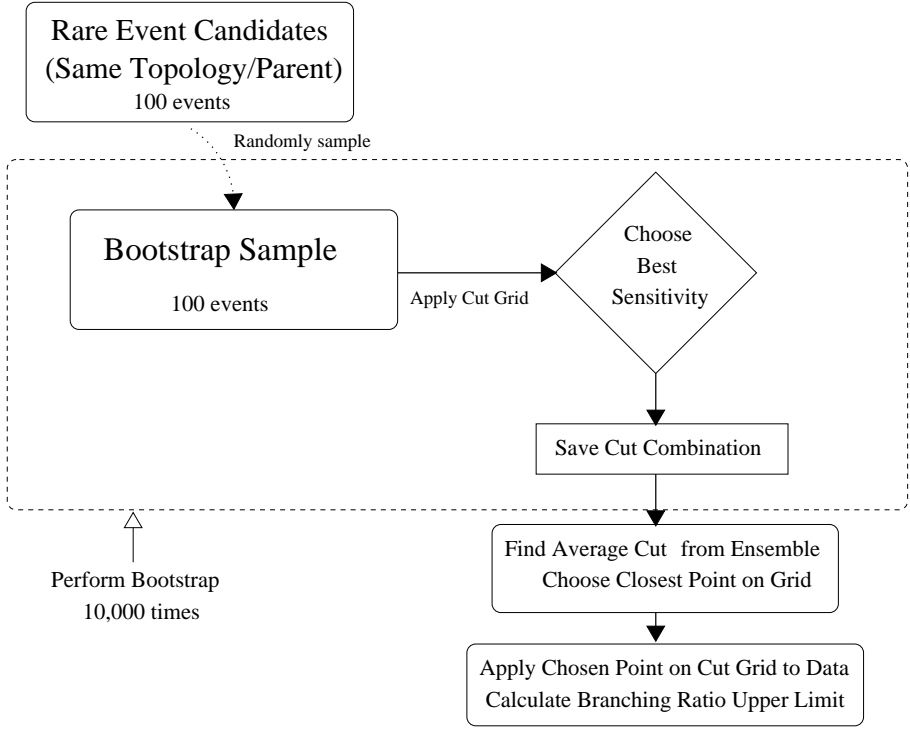,width=12cm}
\caption{A flowchart for the cut bootstrap on a sample size of 100 events. 
This figure should be compared to Figure \ref{fig:bootfig}. In this technique 
we determine an average ``best cut'' for the data using only the data sidebands
and the expected shape of the background.}
\label{fig:cutfig}
\end{center}
\end{figure}

\begin{figure}[htb]
\begin{center}
\center{\epsfig{file=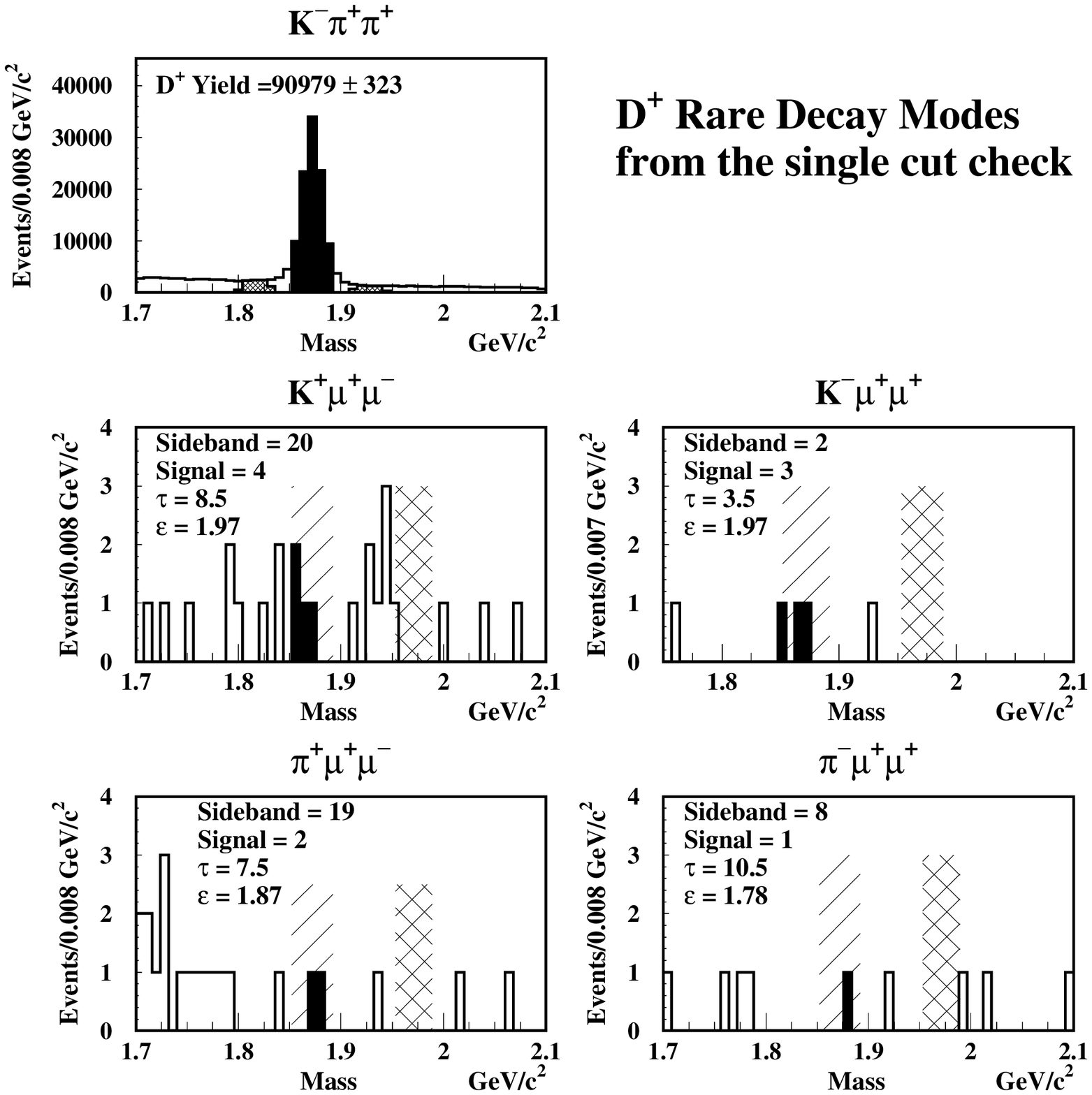,width=14cm}}
\caption{Data used in the single cut systematic check for the $D^+$ decay modes. 
Note that the $\tau$'s shown
on the plots are the same as those used for the Dual Bootstrap analysis. 
The solid histogram entries correspond to events in the signal region.
The cross-hatched areas to either side of the normalization mode signal correspond
to the data used for the sideband subtraction.
The singly hatched areas in the di-muon mode histograms correspond to the
signal region, while the cross-hatched areas correspond to an excluded
region. All the other data and area shown in the di-muon histograms 
are used for the background estimate. }
\label{fig:rar1}
\end{center}
\end{figure}

\begin{figure}[t]
\begin{center}
\epsfig{file=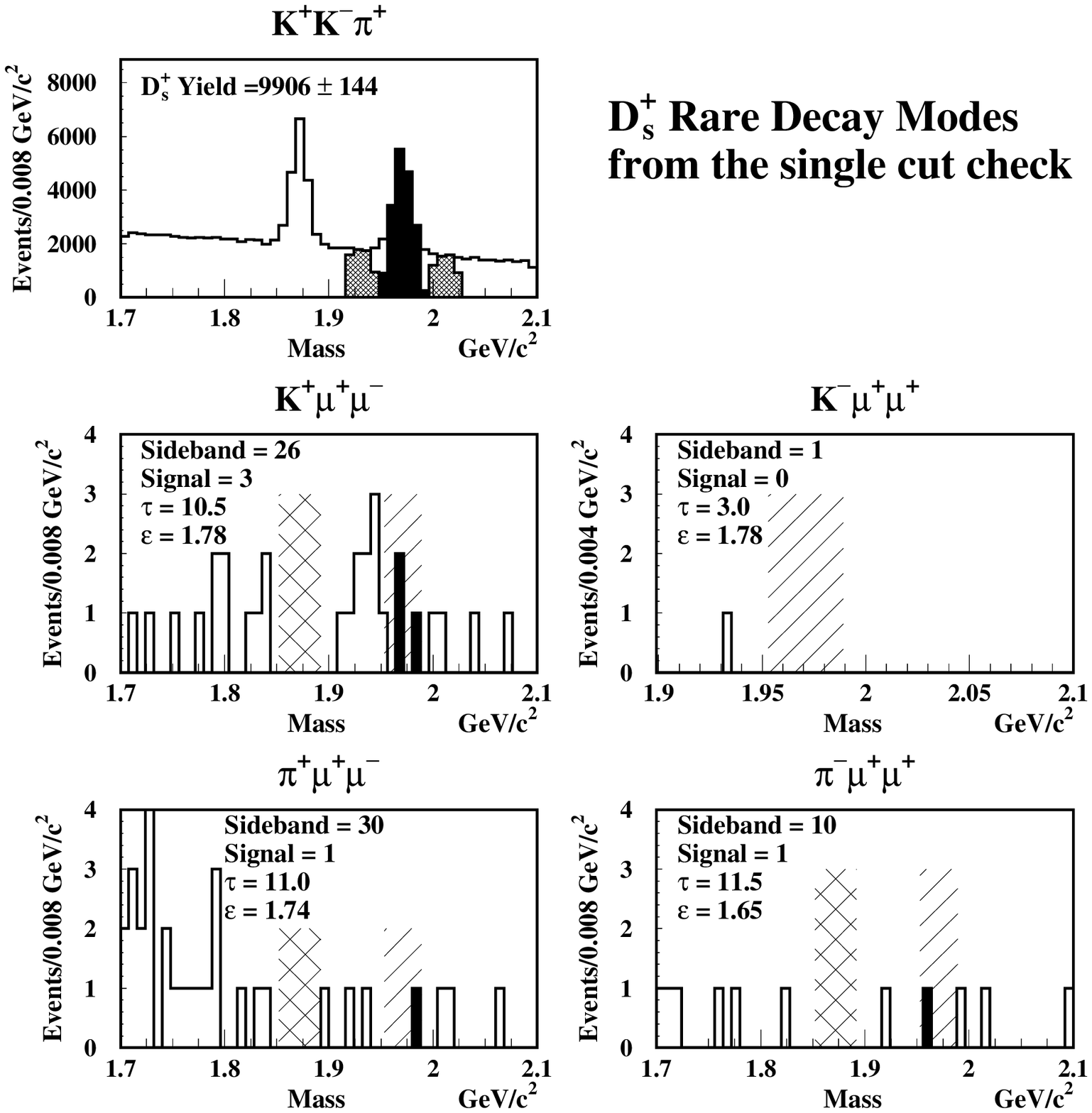,width=14cm}
\caption{Data used in the single cut systematic check for the $D_s^+$ decay modes. 
Note that the $\tau$'s shown
on the plots are the same as those used for the Dual Bootstrap analysis. 
The solid histogram entries correspond to events in the signal region.
The cross-hatched areas to either side of the normalization mode signal correspond
to the data used for the sideband subtraction.
The singly hatched areas in the di-muon mode histograms correspond to the
signal region, while the cross-hatched areas correspond to an excluded
region. All the other data and area shown in the di-muon histograms 
are used for the background estimate.
}
\label{fig:rar2}
\end{center}
\end{figure}

To compare the Dual Bootstrap results to a more traditional ``blind'' analysis, 
another technique was used that selected a unique set of cuts, or point on the 
cut grid. Since the $D^+$ and $D_s^+$ lifetimes and production topology differ,
a separate point on the cut grid was determined for each.
The cuts used to determine the best sensitivities in the first bootstrap
are saved for all four rare modes of a parent particle. A point in the multi-dimensional
cut grid is determined by choosing cuts closest to the average
value of each saved cut (see Figure \ref{fig:cutfig}).
The cuts represented by these 2 cut grid points, one for the $D^+$ and one for the $D_s^+$,  
are then applied to the respective modes {\it{once}}
in the spirit of a more traditional ``blind'' analysis. A branching ratio limit
is computed using the resultant data histogram and the previous definitions
for the signal region, the background sidebands, and $\tau$. 
The best average cuts are shown in Table \ref{tab:cuts}, and
the data histograms resulting
from this check are shown in Figures \ref{fig:rar1} and \ref{fig:rar2}. A 
comparison was
also made between the confidence limit calculated using the Rolke-Lopez method \cite{Rolke:2000ij}
and the Feldman--Cousins method \cite{Feldman:1997qc}. Little difference was seen. 
We stress that these checks are provided as a convenience to the reader. As stated 
previously, the methods of Rolke and Lopez \cite{Rolke:2002ix,Rolke:2000ij} have been demonstrated to 
provide correct coverage, whereas the coverage of the checks mentioned has
either not been studied or, in the case of Feldman-Cousins where
background fluctuations are not considered, has been shown to have incorrect
coverage.


No significant evidence for the observation of any of the rare decay modes was seen.  
All modes except $D^+ \to K^- \mu^+ \mu^+$  had a 90\% lower limit for the branching ratio 
of zero. For $D^+ \to K^- \mu^+ \mu^+ $, the 96\% lower limit was zero. The results of the 
analysis are presented in Table \ref{tab:blood} below. There is 
good agreement between the Dual Bootstrap branching ratios (for both the Feldman-Cousins
and Rolke-Lopez limits), the sensitivities, and the single
cut systematic checks. 

\begin{table}[htb]
\caption{FOCUS results with and without incorporated systematic errors for the modes shown. 
Each number represents a 90\% confidence upper limit for the brancing ratio of the decay mode listed. 
F-C represents the Feldman-Cousins 90\% confidence upper limit. R-L represents the Rolke-Lopez
90\% confidence upper limit. Note the relatively minor differences between the sensitivities,
the Feldman-Cousins limits and the Rolke-Lopez limits.
Our final result is the Rolke-Lopez 90\% confidence upper limit including 
the systematic error shown in the fifth column of the table. The single
cut check result, which also includes the systematic error, shown in the last column of the table,
agrees with our final result as well. All modes are $(\times 10^{-6})$.}
\label{tab:blood}
\begin{center}
\begin{tabular}{ccccccl} \hline 
Decay                     & Sensitivity &F-C    & R-L	& {\bf{R-L incl.}}  & Single Cut   \\
Mode                    & 	    &	    &		& {\bf{$\sigma_r$}} & incl. $\sigma_r$ \\ \hline
$D^+ \to K^+ \mu^+ \mu^- $    & $7.5$	    & $11  $ & $9.1$	  & {\bf{$9.2 $}}   & $12 $       \\ \hline
$D^+ \to K^- \mu^+ \mu^+ $    & $4.8 $      & $13  $ & $13 $	  & {\bf{$13 $}}    & $12 $	     \\ \hline
$D^+ \to \pi^+ \mu^+ \mu^- $  &  $7.6$      & $9.3 $ & $8.7$	  & {\bf{$8.8$}}    &  $7.4 $       	\\ \hline
$D^+ \to \pi^- \mu^+ \mu^+ $  &  $5.5$      & $4.6 $ & $4.8$	  & {\bf{$4.8$}}    &  $5.1 $       	\\ \hline
$D_s^+ \to K^+ \mu^+ \mu^- $  &  $33$	    & $31  $ & $33 $	  & {\bf{$36 $}}    &  $38 $	       \\ \hline
$D_s^+ \to K^- \mu^+ \mu^+ $  &  $21$	    & $11  $ & $13 $	  & {\bf{$13 $}}    &  $20 $	       \\ \hline
$D_s^+ \to \pi^+ \mu^+ \mu^- $&  $31$	    & $20  $ & $24 $	  & {\bf{$26 $}}    &  $18 $	       \\ \hline
$D_s^+ \to \pi^- \mu^+ \mu^+ $&  $23$	    & $29  $ & $26 $	  & {\bf{$29 $}}    &  $22 $	       \\ \hline
\end{tabular}
\end{center}

\end{table}

\begin{table}[b]
\label{tab:res}
\caption{FOCUS results compared to other experiments and recent theory. The previous  
limits, except for the E687 $D^+ \to K^- \mu^+ \mu^+ $ \cite{Frabetti:1997wp} 
are from Fermilab experiment E791 \cite{Aitala:1999db}. The theory
estimates come from \cite{Fajfer:2001sa} (SM-1), \cite{Singer:1996it} (SM-3), and 
\cite{Burdman:2001tf} (R-Parity MSSM and SM-2). 
Note that the SM estimates from \cite{Burdman:2001tf} use a formalism close to
\cite{Singer:1996it}, and at present there is some discrepancy in the invariant
$M_{ll}$ mass behavior for the SM estimates in  SM-3 \cite{Burdman:2001tf} 
and SM-1 \cite{Fajfer:2001sa}. All modes shown are $(\times 10^{-6})$.}

\begin{center}
\begin{tabular}{ccccccccl}
\hline
 Decay  &  This    & SM-1 & SM-2 & SM-3 & MSSM       & Previous \\
 Mode   & Analysis &      &      &      &  R-Parity  &     Best \\
\hline
$D^+   \to K^+\mu^-\mu^+  $ & $9.2$ & $0.007$   &	 -   &	 -   &  -   & $44$    \\
$D^+   \to K^-\mu^+\mu^+  $ & $13$ &   -      &  -    & -   &	 -   & $120$  \\
$D^+   \to \pi^+\mu^-\mu^+$ & $8.8$ & $1.0$   & $1.9$ & $1.8$ & $15$ & $15$   \\
$D^+   \to \pi^-\mu^+\mu^+$ & $4.8$ &   -      & -   &	 -   &  -   & $17$    \\
$D_s^+ \to K^+\mu^-\mu^+  $ & $36$  & $0.043$ &	 -   &	 -   &  -   & $140$   \\
$D_s^+ \to K^-\mu^+\mu^+  $ & $13$  &	 -    &	 -   &	 -   &  -   & $180$   \\
$D_s^+ \to \pi^+\mu^-\mu^+$ & $26$  & $6.1$   &	 -   &	 -   &  -   & $140$   \\
$D_s^+ \to \pi^-\mu^+\mu^+$ & $29$  &	-     &	 -   &	 -   &  -   & $82$    \\
\hline 

\end{tabular}
\end{center}

\end{table}

\section{Summary and Conclusions }

The FOCUS results from this analysis are shown in comparison to previous best results and recent 
theory in Table \ref{tab:res}. Our results are a substantial improvement
over previous results \cite{Frabetti:1997wp,Aitala:1999db} and the FOCUS result for
the branching ratio upper limit $D^+ \to \pi^+ \mu^+ \mu^- $ of $8.8 \times
10^{-6}~\char'100$  90\% C.L. is lower than the current
MSSM R-Parity violating constraint \cite{Burdman:2001tf}
for this mode.

\section*{Acknowledgments}

We wish to acknowledge the assistance of the staffs of Fermi
National Accelerator Laboratory, the INFN of Italy, and the physics
departments of the collaborating institutions. This research was 
supported in part by the U.~S.
National Science Foundation, the U.~S. Department of Energy, the
Italian Istituto Nazionale di F\'{i}sica Nucleare and Ministero
dell'Universit\`a e della Ricerca Scientifica e Tecnologica, 
the Brazilian Conselho Nacional de
Desenvolvimento Cient\'{\i}fico e Tecnol\'ogico, CONACyT-M\'exico,
the Korean Ministry of Education, and the Korean Science and 
Engineering Foundation.
The authors wish to thank Gustavo Burdman and Paul Singer for 
their patience during several very useful conversations.

\end{document}